\documentclass[amsmath,amssymb,aps,twocolumn]{revtex4-2}

\usepackage{graphicx}
\usepackage{dcolumn}
\usepackage{bm}
\usepackage{upgreek}
\usepackage[toc,page]{appendix}

\newcommand{\ket}[1]{{\left| {#1} \right\rangle}}

\newcommand{\ketbra}[2]{{\left| {#1} \right\rangle \!\!\left\langle {#2} \right|}}

\newcommand*{\mathabxbfamily}{\fontencoding{U}\fontfamily{mathb}\selectfont}
\DeclareFontFamily{U}{mathb}{\hyphenchar\font45}
\DeclareFontShape{U}{mathb}{m}{n}{
      <5> <6> <7> <8> <9> <10> gen * mathb
      <10.95> mathb10 <12> <14.4> <17.28> <20.74> <24.88> mathb12
      }{}
\newcommand*{\Sun}{{\text{\mathabxbfamily\char"40}}}

\begin{document}

\preprint{APS/123-QED}

\title{Thermal light in confined dimensions for ``laser'' cooling with unfiltered sunlight}

\author{Amanda Younes}
\affiliation{Department of Physics and Astronomy, University of California Los Angeles, Los Angeles, CA, USA}

\author{Wesley C. Campbell}
\affiliation{Department of Physics and Astronomy, University of California Los Angeles, Los Angeles, CA, USA}

\date{\today}

\begin{abstract}

Cooling of systems to sub-kelvin temperatures is usually done using either a cold bath of particles or spontaneous photon scattering from a laser field; in either case, cooling is driven by interaction with a well-ordered, cold (\textit{i.e.}\ low entropy) system. However, there have recently been several schemes proposed for ``cooling by heating,'' in which raising the temperature of some mode drives the cooling of the desired system faster.  We discuss how to cool a trapped ion to its motional ground state using unfiltered sunlight at $5800\,\mathrm{K}$ to drive the cooling.  We show how to treat the statistics of thermal light in a single-mode fiber for delivery to the ion, and show experimentally how the black-body spectrum is strongly modified by being embedded in quasi-one-dimension.  Quantitative estimates for the achievable cooling rate with our measured fiber-coupled, low-dimensional sunlight show promise for demonstrating this implementation of cooling by heating.

\end{abstract}

\maketitle

\section{Introduction}

Cooling is commonly accomplished by the coupling between a system of interest and a cold bath, with no further interactions.  However, in some cases the coupling between the two is controlled by the occupation of some other mode that connects them.  A good example of this is laser cooling, where the electromagnetic modes populated by laser photons allow the system of interest to repeatedly spontaneously emit into cold vacuum modes \cite{Phillips1998Nobel}.  The cooling is thereby driven by the highly-occupied modes of the laser, without which the cooling rate will fall to essentially zero.

However, since an ideal laser field is often approximated as in a coherent state \cite{Glauber1963Coherent}, it is a displaced vacuum and has no entropy; the laser field can be thought of as being highly-ordered and in that sense also extremely cold \footnote{A comet orbiting the sun can still have a very cold temperature, even if its atoms' motional states have all been displaced to some large, average energy in our reference frame.}.
The lasers used for laser cooling tend to have very narrow linewidths (typically $\Delta \nu/\nu < 10^{-8}$ for laser cooling atoms and molecules), as that feature allows the absorption of laser photons to be velocity-dependent.  One can therefore ask, is it \emph{necessary} for laser cooling that the field that drives the cooling step (\textit{i.e.}\ that couples the system to the cold bath) also be in a low-entropy state?  And if so, to what extent can we identify the highly-ordered nature of \emph{that} laser field, as opposed to the highly-ordered nature of the vacuum field, as being responsible for the cooling?

Here, we propose how the phenomenon known as ``cooling by heating'' \cite{cbh1,cbh2} can be used to illustrate the answer to these questions.  Cooling by heating refers to cases where the coupling between the system of interest and the cold bath can be increased by increasing the thermal occupation of a mode that couples the two, and therefore the system can be cooled by heating that mode.  This paradigm has been used to study some counter-intuitive scenarios exhibiting cooling by heating \cite{mechres,qopt}, and open questions persist about the interplay between this phenomenon and quantum correlations \cite{VillasBoas2016Does} and dissipative generation of entangled states \cite{cbh1}.

We begin by introducing an experimentally-accessible scenario from atomic physics where the motion of a single, trapped atomic ion is to be cooled to its quantum ground state via a repeated cycle that uses sunlight to drive the coupling to the cold vacuum.  Unitary (and therefore reversible, entropy-conserving, and non-cooling) operations on the atom's state are driven by a laser, followed by a separate step where only sunlight is applied to cool the ion's temperature.  We analyze the achievable temperature in the presence of multiple baths at different temperatures using a model of virtual qubits \cite{virtual,Mitchison2016Realising}.  We then discuss the statistical physics of black-body radiation confined in a single-mode optical fiber for delivery to the ion, and observe how dimensionality affects the spectrum of a black body by analyzing fiber-coupled sunlight with a spectrometer.  We conclude with an estimate of the achievable experimental cooling rate in this system.

\section{Cooling an ion with thermal light}

The form of laser cooling that we will consider for this demonstration is known as resolved-sideband cooling \cite{Wineland1975Proposed,Dehmelt1976Entropy}, and has been implemented with lasers to cool ions \cite{Diedrich1989Laser}, atoms \cite{Hamann1998Resolved}, and micromechanical oscillators \cite{Teufel2011Sideband} to their quantum ground states of motion.

For a harmonically trapped atomic ion, the ion's motion in the trap is periodic at frequency $\omega_\mathrm{motion}$ and this gives rise to the appearance of phase-modulated sidebands on the spectrum of applied laser light as observed in the rest frame of the ion.  In the lab frame, this means that the ion's optical absorption spectrum consists not only of a ``carrier'' peak at the rest-frame resonant frequency of some optical transition (call it $\omega_1$), but also sidebands at $\pm \omega_\mathrm{motion}$ from the carrier (as well as at $\pm 2 \omega_\mathrm{motion}$, and so on).  If the laser frequency ($\omega_\ell$) is set to be resonant with the feature at $\omega_\ell = \omega_1 - \omega_\mathrm{motion}$ (the ``red sideband''), the ion can absorb photons with energy $\hbar \omega_\ell$ but it will emit them with an average energy closer to $\hbar\omega_1$.  Each cycle, then, removes approximately $\hbar \omega_\mathrm{motion}$ of thermal energy on average, cooling the ion.  Since the strength of the red sideband goes to zero as the ion's motion approaches the ground state, resolved-sideband cooling is capable of cooling the ion to its quantum ground state and then ceases to have any effect (in the absence of other sources of heating).

Since the spatial extent of the ion's harmonic motion ($x_0 = \sqrt{\hbar/(2 m \omega_\mathrm{motion})}$, typically $\approx 5 \,\mathrm{nm}$) is much smaller than the wavelength of radiation at that frequency ($\lambda = 2 \pi c/\omega_\mathrm{motion}$, typically $\approx 1\, \mathrm{km}$), it is a very poor antenna, and the ion's motion is substantially impedance mismatched to electromagnetic radiation.  In the absence of technical electric field noise, the motion of ions trapped inside room-temperature vacuum chambers remain out-of-equilibrium with the chamber for all relevant experimental timescales, and we will ignore any direct coupling between light and motion.

\begin{figure}[t]
    \centering
    \includegraphics[width = 0.45\textwidth]{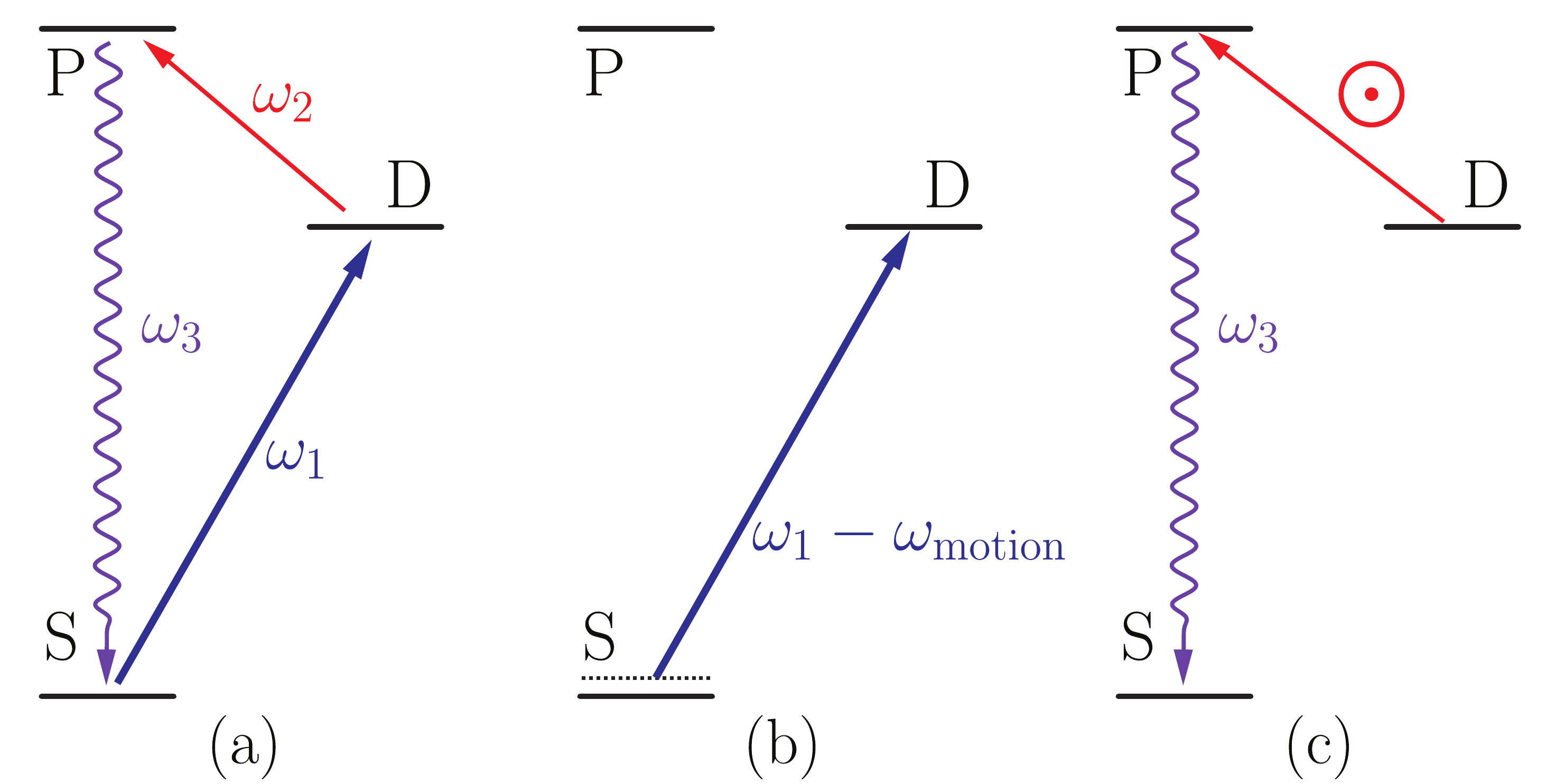}
    \caption{Atomic level structure for sideband cooling. (a) A three-level atom with an $\mathrm{S} \leftrightarrow \mathrm{D}$ transition at $\omega_1$, a $\mathrm{D} \leftrightarrow \mathrm{P}$ transition at $\omega_2$, and a $\mathrm{S} \leftrightarrow \mathrm{P}$ transition at $\omega_3$.  (b) Step I of the cooling cycle, in which a red sideband of the $\mathrm{S} \leftrightarrow \mathrm{D}$ transition is driven by a laser at frequency $\omega_\ell = \omega_1 - \omega_\mathrm{motion}$. (c) Step II of the cooling cycle, in which any population in $\mathrm{D}$ is returned to $\mathrm{S}$ via excitation to $\mathrm{P}$ from absorption of a photon at $\omega_2$ followed by spontaneously emitting a photon at $\omega_3$.  We propose that the light at $\omega_2$ can be provided by black-body radiation from the sun.}
    \label{fig:AtomicLevels}
\end{figure}

The scheme we consider for sideband cooling of a trapped ion is shown in figure \ref{fig:AtomicLevels}, and consists of a repeated, two-step cycle as follows.  In step I (Fig.~\ref{fig:AtomicLevels}(b)), the ion is illuminated by a narrow-linewidth laser on the red sideband of the $\mathrm{S} \rightarrow \mathrm{D}$ transition.  The intensity and illumination time are chosen to fully transfer population to the (long-lived) $\mathrm{D}$ state for ions with energy near their thermal average energy, and then the laser is turned off.  On average, this step can reduce the motional energy of the ion, but it also adds a much larger amount of total energy in the form of internal excitation. Entropy from the motional state has been partially transferred to the internal state of the ion in this unitary process. Since step I is reversible, the total entropy of the ion has not changed and the laser has accomplished no cooling.  

In step II (Fig.~\ref{fig:AtomicLevels}(c)), the ion is illuminated with light capable of driving any population in the long-lived $\mathrm{D}$ state to a higher-lying $\mathrm{P}$ state (with resonant frequency $\omega_2$) that can quickly decay to the ground $\mathrm{S}$ state by spontaneously emitting a photon ($\omega_3$) into an approximately unoccupied mode (thermal states of optical-frequency modes at room temperature are close to the vacuum state).  This step does not change the motional energy of the ion on average, but does reduce the total energy since the ion returns to its internal ground state.  However, unlike step I, this step is not reversible and the ion's entropy (and therefore temperature) has been reduced.  The optical modes at frequency $\omega_3$ contain information about the ion's motional state, which is to say that the spontaneously emitted photon carries away entropy.  This is the cooling step in the process.

In a typical implementation of sideband cooling for applications in precision measurement or quantum information processing, step II is driven by a laser at $\omega_2$.  However, this light need not be coherent nor narrow in linewidth, as its only job is to couple the ion to the vacuum modes at $\omega_3$.  For this, we propose to use thermal light in the form of fiber-coupled black-body radiation at $T_\Sun \approx 5800 \, \mathrm{K}$ from the sun.  Even if this light has spectral density near $\omega_1$ and $\omega_3$, the driving of those transitions will not directly change the motional energy of the atom on average, and this does not inhibit cooling to the ground state (see Appendix \ref{appendix:AllThermal}).

\subsection{Minimum Achievable Temperature}
To estimate the minimum temperature that can be achieved with this scheme, we start with a steady-state version of the sequence shown in Fig.~\ref{fig:AtomicLevels}(b) and (c) to illustrate thermalization, for which we assume band-limited sunlight and utilize the concept of virtual qubits \cite{virtual}.  Following that, we analyze the time-dependent scheme with unfiltered sunlight and argue that it achieves the same limiting temperature insofar as both scenarios yield a predicted minimum temperature that is below what the limiting temperature will likely be in practice due to other heating effects (such as momentum diffusion from absorption and emission, or heating from electrical noise).

If we wish to apply principles of thermodynamic equilibrium to the trapped ion example in Fig.~\ref{fig:AtomicLevels}, we need to identify a bath that is brought into contact with the ion's motion to cool it.  For this, we can simplify the time-dependent scheme into a steady-state scheme by assuming that both the narrow-band laser light (Fig.~\ref{fig:AtomicLevels}(b)) and the sunlight ($T_2 = T_\Sun$) that connects levels D and P (Fig.~\ref{fig:AtomicLevels}(c)) are applied simultaneously and continuously, and that the sunlight is band-limited such that it does not connect either upper state to S.  In this case, the $\mathrm{S} \leftrightarrow \mathrm{P}$ transition is driven by room-temperature ($T_3 = T_\mathrm{room}$) black-body radiation, and we can model the effect of the laser on $\mathrm{S} \leftrightarrow \mathrm{D}$ as a thermal field with temperature $T_\ell \rightarrow \infty$.

The continuous interaction of these three subsystems, each with a unique temperature, with the ion's motion can be aggregated into an effective interaction of the motion with a single virtual qubit with splitting $\omega_\mathrm{V} = \omega_\mathrm{motion}$ held at a virtual temperature $T_\mathrm{V}$, which is given by \cite{virtual} (also, Appendix \ref{appendix:VirtualQubits}):
\begin{align}
    T_\mathrm{V} =& \,\, \frac{\omega_\mathrm{V}}{ \frac{ \omega_3}{T_3} - \frac{\omega_2}{T_2} - \frac{ \omega_\ell}{T_\ell} } \nonumber \\
    = & \,\,\frac{\omega_\mathrm{motion}}{ \frac{\omega_3}{ T_\mathrm{room}}  -\frac{ \omega_2}{T_\Sun} }.
\end{align}
In the limit that $\omega_3/T_\mathrm{room} \gg \omega_2/T_\Sun$, we conclude that the minimum achievable temperature is given by the virtual temperature
\begin{equation}
    T_\mathrm{V} \approx \frac{\omega_\mathrm{motion}}{\omega_3} T_\mathrm{room}. \label{eq:ScaledTroom}
\end{equation}
A thermal state of motion at $T_\mathrm{V}$ has an average motional excitation of $\langle n_\mathrm{motion} \rangle = \left(\exp \left( \beta_\mathrm{V} \hbar \omega_\mathrm{motion}  \right) - 1 \right)^{-1} \approx \exp \left(-\beta_\mathrm{V} \hbar \omega_\mathrm{motion}\right)$ where $\beta_i^{-1} \equiv k_\mathrm{B} T_i$.  Since we expect $k_\mathrm{B} T_\mathrm{room} \ll \hbar \omega_3$, this corresponds to an ion in its ground state of motion ($T_\mathrm{V} \approx 1\,\upmu \mathrm{K}$ and $\bar{n}_\mathrm{motion} \approx 10^{-46}$).  Even in the limit where $T_\mathrm{room}$ is replaced with $T_\Sun$, the virtual temperature is cold enough to cool the ion to its ground state (see Appendix \ref{appendix:AllThermal}).  We stress here that there are many potential sources of heating in experiments that we have not attempted to capture with this analysis, which focuses only on the steady state solution with thermal radiation fields (the ``black-body limit'').  Our conclusion is that ground-state cooling is possible with this scheme, but we do not claim that it is necessarily practical to achieve cooling all the way to $T_\mathrm{V}$.

For the time-dependent scheme of Fig.~\ref{fig:AtomicLevels}(b) and (c) with unfiltered sunlight, if the sunlight is left on long enough for the atom's internal states to equilibrate, the atomic energy distribution will be held at $T_\Sun$.  However, the periodic extinction of that light in Step I, and in particular the case if we assume that the light is turned off at the end of the protocol and the ion's internal states are allowed to relax, indicates that the atomic internal temperature will equilibrate to $T_\mathrm{room}$.  As such, the effect of the laser is essentially to translate the motional thermal state to an energy scale set by $\omega_1$ and allow that system to thermalize to $T_\mathrm{room}$, followed by relaxation of that atomic excitation back to the energy scale of $\omega_\mathrm{motion}$.  We therefore expect the motional temperature to be limited by these considerations to a scaled version of $T_\mathrm{room}$, which is precisely the temperature in Eq.~(\ref{eq:ScaledTroom}).  This is also the black-body limit for standard sideband cooling with laser light in a room-temperature vacuum chamber.

\subsection{Excitation Rate from fiber-coupled black-body radiation}
An atom illuminated by a focused beam of thermal light will not experience the same field as if it were inside a black body.  Many modes will be in vacuum (or at least different thermal) states due to the anisotropy of the illumination, rather than in thermal states at the temperature of the black body.  To calculate the excitation rate, $\Gamma$, of a two-level atom initially in its ground state (state manifolds $\{ \ket{\mathrm{e}_i} \}$ and $\{ \ket{\mathrm{g}_i} \}$ with degeneracies $g_\mathrm{e}$ and $g_\mathrm{g}$, split by $\omega_\mathrm{eg}$) illuminated by focused incoherent light, we adopt an Einstein rate equation approach,
\begin{align}
    \Gamma = & \,\, B_\mathrm{ge} \rho(\omega_\mathrm{eg}) \\
    = & \,\, \frac{\pi^2 c^3}{\hbar \omega_\mathrm{eg}^3} \frac{g_\mathrm{e}}{g_\mathrm{g}} A_\mathrm{eg}\, \rho(\omega_\mathrm{eg}) \label{eq:WCCRate}
\end{align}
where $A_\mathrm{eg}$ and $B_\mathrm{ge}$ are the Einstein A and B coefficients for the transition and $\rho(\omega_\mathrm{eg})$ is the spectral energy density (energy per unit volume per unit angular frequency) at the ion's position at the transition frequency.

We consider that light from an ideal black body is coupled into an optical fiber with a single, gaussian transverse mode and this fiber's output is being imaged onto the atom with an imaging system having half-cone convergence angle $\vartheta$.  We will further suppose, to keep our analysis consistent with typical experimental hardware, that $\vartheta \ll 1$, which allows us to treat the optical system with the paraxial approximation.  We will assume that the fiber-coupled thermal light is the only significant source of illumination to calculate the rate from that contribution alone.

In the typical, textbook treatment of black body radiation, the light inside a black body in equilibrium at temperature $T$ is characterized by a constant, isotropic spectral radiance  $B(\omega)$ (power per unit solid angle, per unit area, per unit angular frequency) given by Planck's law of black-body radiation, 
\begin{align}
B_\mathrm{P}(\omega) = \frac{\frac{\hbar \omega^3}{4 \pi^3 c^2}}{\exp \left( \beta \hbar \omega\right) - 1}.\label{eq:PlancksBP}
\end{align}
However, as we show in the next section, thermal light emerging from a single-mode fiber differs somewhat from a true black body, and is more conveniently characterized instead by a power spectral density $S(\omega)$ (power per unit angular frequency of a single transverse spatial mode).  If the imaging system is capable of focusing the fiber mode onto a (potentially frequency-dependent) effective mode area $A(\omega)$, the spectral energy density at the atom is given by
\begin{align}
    \rho(\omega) = & \,\, \frac{S(\omega)}{c A(\omega)},\label{eq:rho}
\end{align}
which can be used with Eq.~(\ref{eq:WCCRate}) to calculate the excitation rate.

Given an instantaneous excitation rate, $\Gamma p_\mathrm{D}$, from state D to state P (see Fig.~\ref{fig:AtomicLevels}) for an atom with probability $p_\mathrm{D}$ of being in state D, the rate at which this results in a spontaneous emission back to the ground state, which completes the step of removing one quantum of motion on average, will be
\begin{equation}
    \dot{n}_\mathrm{motion} = -\Gamma\, p_\mathrm{D}\, \eta_\mathrm{SP}
\end{equation}
where $\eta_\mathrm{SP} \equiv A_\mathrm{PS}/(A_\mathrm{PS} + A_\mathrm{PD})$ is the branching fraction of spontaneous emission from P to go back to S.

\section{Thermal light in a single-mode fiber}

The statistics of thermal light confined to quasi-one-dimension (q1D \footnote{We are deliberate in not to referring to this as ``one-dimensional light,'' which is a different topic entirely.  Maxwell's equations treating the light in the fiber are fully three-dimensional, and the embedding in quasi-one-dimension refers to the fact that there is only one mode available in each of the two transverse directions.}) has been discussed in various contexts, including Johnson-Nyquist noise \cite{Nyquist1928Thermal,Oliver1965Thermal}, photonics \cite{Fohrmann2015Single,Fisenko2019BlackBody}, photovoltaic energy conversion \cite{DeVos1988Thermodynamics}, and extra spatial dimensions \cite{Landsberg1989StefanBoltzmann,Alnes2007Blackbody}.  We do not, therefore, present a new theoretical result by deriving the power spectral density of black-body radiation embedded in q1D.  Here, we present an optics-oriented derivation to illustrate the origin of the spectrum we will use to compare to experimental observations, and a brief discussion of how to reconcile the modified spectrum in the fiber with Planck's law.

We can consider a single-mode optical fiber (which is to say, some waveguide that only supports one transverse mode of the electromagnetic field at each frequency $\omega$) of length $L$ (later, we will take $L \rightarrow \infty$) with periodic boundary conditions and light allowed to propagate in only one of the two possible directions.  Assuming, for simplicity, that the effective index of refraction in the fiber is $n=1$, the allowed frequencies will be $\omega_i = i \times 2 \pi c/L$, and the density of states per polarization will therefore be $\mathrm{d}i/\mathrm{d}\omega = L/(2 \pi c)$.

The average rate of photons in mode $i$ passing through a fixed reference plane in the fiber will be $\langle n_i \rangle c/L$, so the time-averaged power from mode $i$ is $P_i = \hbar \omega_i \langle n_i \rangle c / L$, where $\langle n_i \rangle$ is the average number of photons in mode $i$.  Summing over the two available polarizations, using  $\langle n_i \rangle = \left(\exp \left( \beta \hbar \omega_i  \right) - 1 \right)^{-1} $ for the expected thermal population for a mode with splitting $\hbar \omega_i$ and temperature $T = 1/(k_\mathrm{B} \beta)$, and taking the $L \rightarrow \infty$ limit, we have the total, time-averaged power
\begin{align}
    P_\mathrm{total} 
  = \int_0^\infty \! \mathrm{d} \omega \frac{ \frac{\hbar \omega}{\pi}}{\exp \left( \beta \hbar \omega \right) - 1} = \frac{\pi}{6 \hbar} \frac{1}{\beta^2}. \label{eq:Ptotal}
\end{align}
From the integrand, we identify the power spectral density for thermal light in a single-mode fiber,
\begin{align}
    S(\omega) =  \frac{ \frac{\hbar \omega}{\pi}}{\exp \left( \beta \hbar \omega\right) - 1}.\label{eq:S(omega)}
\end{align}
This expression was used by Nyquist in 1928 to explain thermal noise in electrical circuits \cite{Nyquist1928Thermal}, but it is also the spectrum of power for thermal light coupled into a single-mode fiber.

The spectrum of $S(\omega)$ ($\propto \omega \langle n \rangle$) differs in shape from the spectral radiance given by Planck  ($B_\mathrm{P}(\omega) \propto \omega^3 \langle n \rangle$), and the total power is proportional to $T^2$, as opposed to the more-familiar $T^4$ of the Stefan-Boltzmann law in three dimensions. Thermodynamics, however, requires that the two ends of the fiber, if brought into optical contact with two isolated black bodies, will allow them to equilibrate through the fiber.  How this is possible if the spectrum in the fiber has a different shape and peak position than the three-dimensional case can be resolved as follows.  We consider two extremes for the transverse mode confinement in the fiber: (i) the \emph{divergence angle} of light from the fiber end is independent of $\omega$, which is approximately true for a total-internal-reflection interpretation of step-index fiber; and (ii) the \emph{mode area} of the fiber is independent of $\omega$, which is approximately true for a photonic crystal fiber.  Cases that are intermediate between these two are likewise handled as follows.

For case (i), optical considerations dictate that the effective mode area, $A(\omega)$, must be frequency-dependent to maintain a frequency-independent solid angle $\Omega$.  For case (ii), $\Omega(\omega)$ must be frequency dependent to ensure that $A$ is independent of frequency.  For cases between these two extremes, the relationship between the area of a diffraction-limited mode and its solid angle are related by a well-known phase-space-volume theorem in antenna theory, namely that their product must be equal to the square of the wavelength:
\begin{equation}
    A(\omega) \Omega(\omega) = \lambda^2 =\left(\frac{2 \pi c}{\omega} \right)^2. \label{eq:lambdasquared}
\end{equation}
As pointed out by Dicke in 1946 in the context of thermal noise in microwave systems \cite{Dicke1946Measurement}, this connects the 1D power spectral density to the radiance, 
\begin{align}
B(\omega) = \frac{S(\omega)}{A(\omega) \Omega(\omega)}. \label{eq:BSAomega}
\end{align}
Since thermodynamics requires that this be equal to Eq.~(\ref{eq:PlancksBP}) in thermal equilibrium, this argument highlights that Eq.~(\ref{eq:lambdasquared}) is a basic consequence of Planck's law.  The assignment of spectral radiance for single spatial modes is discussed in Appendix \ref{appendix:SRSM}.

Earlier, we argued that the spectral energy density ($\rho(\omega)$) at the center of the focus of an optical system imaging a single mode of thermal radiation onto a spot size $A(\omega)$ was given by Eq.~(\ref{eq:rho}).  Since the spectral radiance of that light will have the same frequency dependence (spectrum) as Planck's law (\ref{eq:PlancksBP}) but lower power, the thermal light can be called gray-body radiation, and we can use the ratio of the energy spectral density to that inside an ideal black body to define an efficiency (or geometric grayness) factor for the thermal light delivery system,
\begin{align}
    G \equiv \frac{\rho(\omega)}{\rho_\mathrm{P}(\omega)}
\end{align}
where
\begin{align}
    \rho_\mathrm{P}(\omega) = & \,\,  \frac{\omega^2}{\pi c^3} S(\omega) =
    \frac{ \frac{\hbar \omega^3}{\pi^2 c^3}}{\exp \left( \beta \hbar \omega\right) - 1} \label{eq:rhoP}
\end{align}
is the energy density inside an ideal black body.

Combining (\ref{eq:rho}) with (\ref{eq:lambdasquared}) and (\ref{eq:rhoP}) allows us to write the geometric grayness as
\begin{align}
    G = \frac{\frac{\lambda^2}{4 \pi}}{A(\omega)} = \frac{\Omega(\omega)}{4 \pi}
\end{align}
$\Omega(\omega)$ is the solid angle of the mode of the imaging system.  In the limit that the mode solid angle covers all of the available solid angle, we recover the ideal black-body energy density and $G \rightarrow 1$.

\section{Measured power spectrum of fiber coupled sunlight}

To observe the predicted spectrum and power spectral density of Eq.~(\ref{eq:S(omega)}) and benchmark the optical power that can be coupled onto a trapped ion, three fibers with distinct guiding regimes were employed: a single-mode step index fiber, a single-mode photonic crystal fiber, and a step-index multi-mode fiber.  The first two were discussed above, and the multi-mode fiber was used to compare to the 3D spectrum.   

For each fiber, sunlight was coupled in using a roof-mounted, home-built sun tracker and a commercial, aspheric fiber collimator lens as the collection optic.  To keep sunlight maximally coupled, the collimator lens is much larger than the minimum diameter necessary to resolve the sun from a point source (which would be a diameter of $D_\mathrm{min} \approx 100 \,\upmu \mathrm{m}$), allowing for steady coupling efficiency even with pointing instability.  This system is able to maintain maximal coupling for many hours.

To estimate the effect of atmospheric absorption and scattering, as well as non-ideal emission, we use a standard reference spectrum \cite{refspec} for sunlight on the surface of the earth for the case of a collecting lens oriented toward the sun. We use the ratio of the ideal 3D Planck spectrum to the standard spectrum to create the expected standard spectrum in q1D.
This spectrum is an average correction for the sun at a specific elevation in specific atmospheric conditions and does not apply perfectly to our conditions at each measurement; the true correction varies somewhat depending on the elevation of the sun, weather conditions, and pollution levels.

To measure the power spectrum of the fiber-coupled light, we measured the output with a fiber-coupled spectrometer \footnote{Thorlabs CCS175}.  We correct the measured output with the response function provided by the manufacturer. For the single mode fibers, we also correct for the wavelength dependence of light entering the spectrometer through a slit using the mode properties in the fiber specifications. This correction is done by treating the fiber mode as a gaussian beam between the fiber tip and the slit, then cutting off parts of the beam that are blocked by the slit.

\begin{figure}[t]
    \centering
    \includegraphics[width=.8\columnwidth]{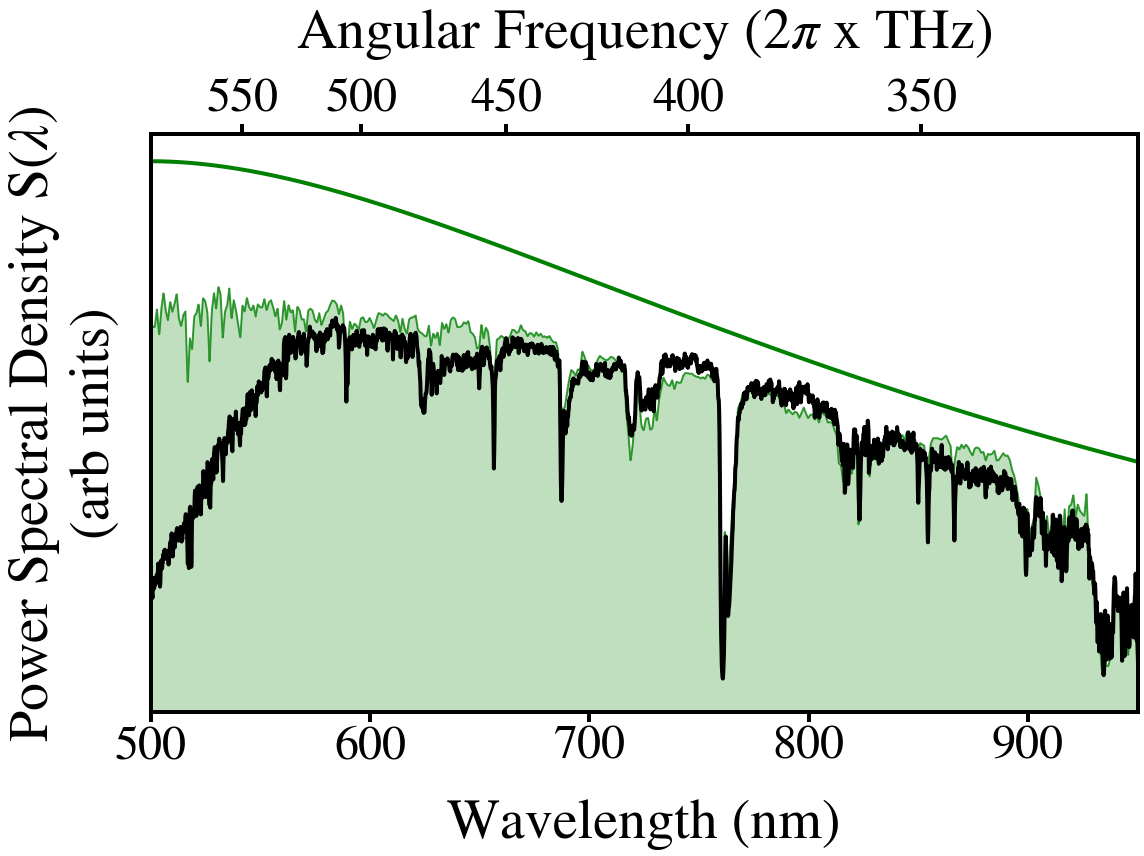}
    \caption{Measured spectrum of sunlight coupled into a step-index, multi-mode optical fiber (black).  The green trace shows the theoretical frequency-dependence of light emitted by an ideal 3D black body, normalized to a peak height near the measured value to allow comparison.  The shaded green-line trace shows the theoretical spectrum, with the atmospheric correction.}
    \label{fig:mm}
\end{figure}

For the multi-mode fiber, Figure \ref{fig:mm} shows that we observe a frequency dependence similar to the standard 3D Planck spectrum, Eq.~(\ref{eq:rhoP}), since the fiber-coupled light can occupy many transverse modes.  The vertical scale in this case is arbitrary, and we have roughly matched the height of the measured and predicted spectra to allow comparison of their shapes.

\begin{figure}[t]
    \centering
    \includegraphics[width=.8\columnwidth]{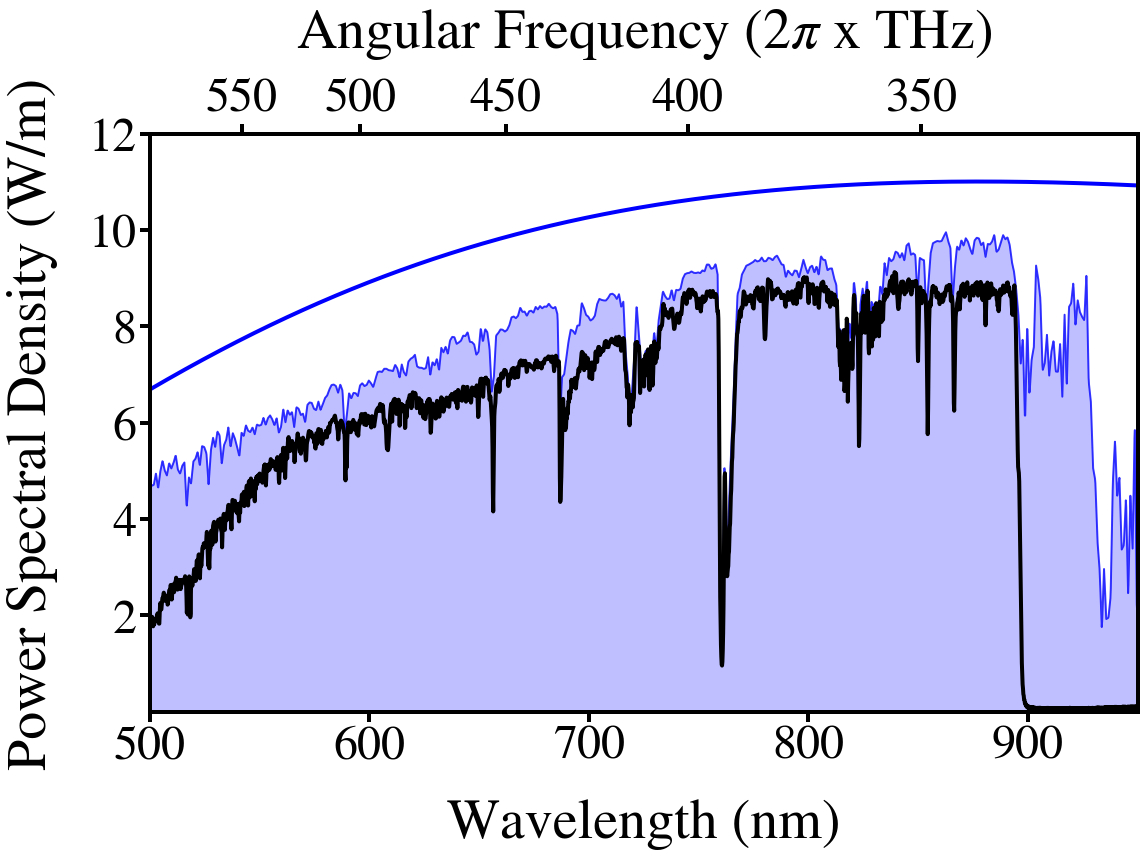}
    \caption{Measured power spectral density of sunlight coupled into a single-mode step-index optical fiber (black), with wavelengths longer than $900 \,\text{nm}$ filtered out. The measured spectrum agrees with the theoretical curve for an ideal black body in q1D (blue, Eq.~(\ref{eq:S(omega)} and the spectrum with atmospheric correction for the q1D case (blue, shaded).}
    \label{fig:step_index}
\end{figure}

\begin{figure}[t]
    \centering
    \includegraphics[width=.8\columnwidth]{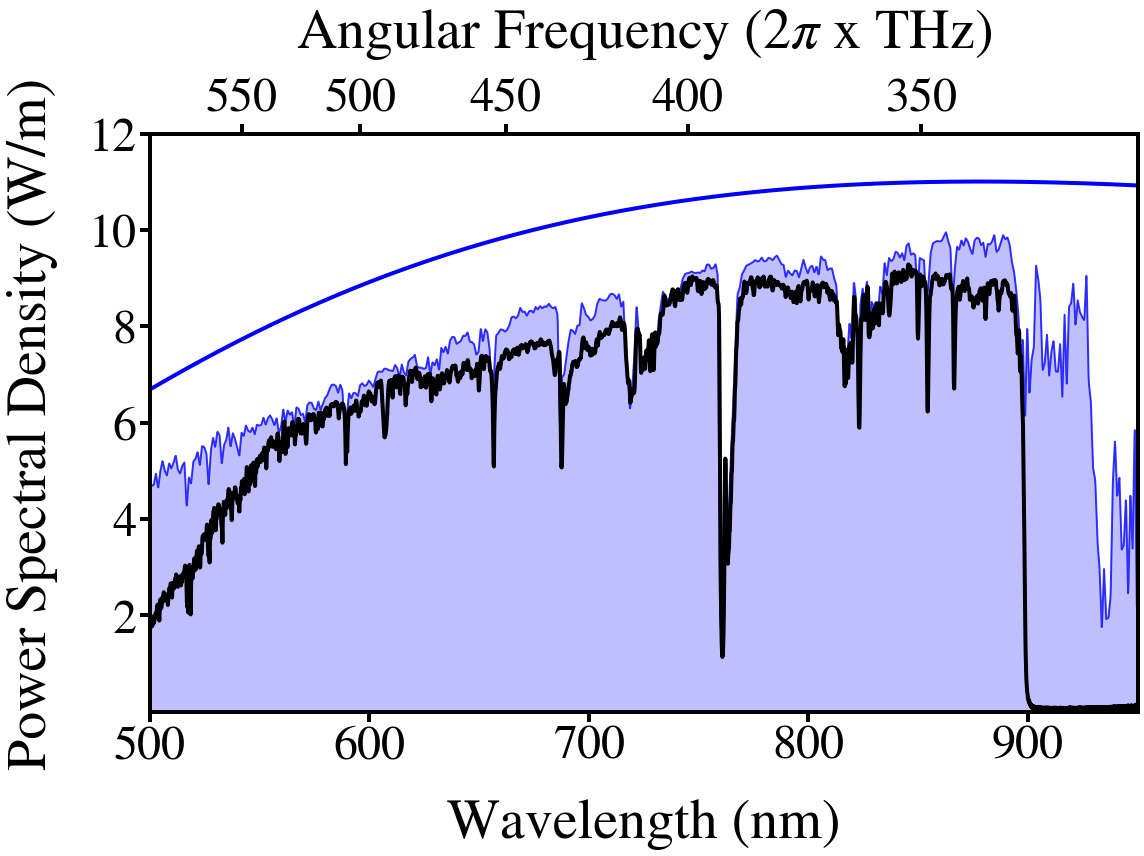}
    \caption{Measured power spectral density of sunlight coupled into a photonic crystal fiber (black), with wavelengths longer than $900 \,\text{nm}$ filtered out. The blue and blue shaded traces are the single mode spectrum and the spectrum with atmospheric correction, respectively, as in Fig.~\ref{fig:step_index}.
    }\label{fig:photonic_crystal}
\end{figure}

For the two single-mode fibers, the measured spectra are shown in black in figures \ref{fig:step_index} and \ref{fig:photonic_crystal}.  The shapes of these spectra are clearly modified from Planck's law in 3D (compare to Fig.~\ref{fig:mm}).  The predicted spectrum of Eq.~(\ref{eq:S(omega)}) is shown in dark blue, along with a prediction that takes into account the empirical solar spectrum at the Earth's surface (shaded blue).

To calibrate the vertical scale, a short-pass filter is inserted in front of the input collimator to remove light with wavelengths longer than $\lambda = 900 \, \text{nm}$, and the power delivered by the fiber is measured with a calibrated photodiode power meter.  By matching this power to numerical integration of the measured spectrum, we obtain power spectral density.

By comparing this to Eq.~(\ref{eq:S(omega)}), we obtain the delivery efficiency, $\eta(\omega)$, as the ratio of our measured power spectral density to the ideal q1D spectrum at $T_\Sun$.  In the visible and near-infrared, we find efficiencies of $\eta = 0.6 - 0.9$ under good seeing conditions with both types of single-mode fiber.

\section{Estimate of cooling rate from measured fiber output}

We can now estimate the achievable cooling rate using fiber-coupled sunlight for our experimental parameters as follows.
We consider a demonstration with $\mathrm{Ba}^+$, for which the cooling light ($\omega_2$ in figure \ref{fig:AtomicLevels}) is near a wavelength of $614\,\text{nm}$. 

Loss in $200 \, \text{m}$ of optical fiber from the roof to the lab is measured to be $35\% - 40\%$ at this wavelength, giving an overall delivery efficiency of $\eta \approx .35 - .60$.  For estimates of cooling rates, we will use $\eta \approx .50$, which we have measured in the lab.

The imaging system is capable of focusing the fiber output to a spot size of $w_0 = 20 \, \upmu \text{m}$, which corresponds to a geometric grayness of $G = 5 \times 10^{-5}$.  The power spectral density in this mode can be related to Eq.~(\ref{eq:S(omega)}) via
\begin{align}
    S_\mathrm{exp}(\omega_2) = \eta \, G \, S(\omega_2) \approx 2.5\times 10^{-5}\,S(\omega_2).
\end{align}

Combining this power spectral density with the atomic parameters for $\mathrm{Ba}^+$ yields an expected cooling rate of 
\begin{align}
    \dot{n}_\mathrm{motion} = -\text{8.2} \text{ phonon}/\text{s}  
\end{align}
for initial motional states well above the ground state.
Heating rates that are comparable (or smaller) in magnitude to this cooling rate estimate ($\dot{n}_\mathrm{motion} < 10\,\text{phonon}/\text{s}$) have been measured for ions in Paul traps \cite{Roos1999Quantum,Daniilidis2014Surface,Randall2016HighFidelity,Bruzewicz2015Measurement,Li2022Robust,Kalincev2021Motional}, and lower rates ($\dot{n}_\mathrm{motion} < 1\,\text{phonon}/\text{s}$) have been measured in in Penning traps \cite{Goodwin2016ResolvedSideband,Stutter2018Sideband}.  It therefore appears feasible that fiber-coupled sunlight may be capable of cooling a trapped ion to its ground state of motion.

\begin{acknowledgments}
The authors acknowledge Kristian Barajas for discussions.
\end{acknowledgments}

\bibliography{Sunlight}

\appendix
\begin{appendices}
\section{Derivation of Virtual Qubit Temperature}\label{appendix:VirtualQubits}

We consider the total Hilbert space of our system to be composed of a subspace for the motion of the ion, $\mathcal{H}_\mathrm{motion}$, with Hamiltonian $H_\mathrm{motion} = \hbar \omega_\mathrm{motion}(a^\dagger_\mathrm{motion}a_\mathrm{motion} + \frac{1}{2}) $; the laser field $\mathcal{H}_\ell$ with Hamiltonian $H_\ell = \hbar \omega_\ell(a^\dagger_\ell a_\ell + \frac{1}{2})$; the field of the sunlight in the vicinity of $\omega_2$, $\mathcal{H}_\Sun$, with Hamiltonian $H_\Sun = \hbar \omega_2(a^\dagger_\Sun a_\Sun + \frac{1}{2})$; and the atomic excitations $\mathcal{H}_\mathrm{atom}$ with Hamiltonian $H_\mathrm{atom} = \hbar \omega_3 \ketbra{\mathrm{P}}{\mathrm{P}} + \hbar \omega_1 \ketbra{\mathrm{D}}{\mathrm{D}}$.  For a single cycle of the cooling scheme shown in Fig.~\ref{fig:AtomicLevels}, we need only consider a 2-level subspace of the optical fields (which we will denote with primes, as in $\mathcal{H}'$), and we can consider them in the Fock basis as consisting of $\ket{n}$ and $\ket{n-1}$ photons for the state before and after an absorption event, respectively.  For the atomic subspace, we can proceed by considering only the 2-level system spanned by $\ket{\mathrm{S}}$ and $\ket{\mathrm{P}}$, and the Hamiltonian will be $H_\mathrm{atom}^\prime = \hbar \frac{\omega_3}{2} \sigma_Z$ where we will refer to this two-level system with spin notation: $\sigma_Z \equiv \ketbra{\mathrm{P}}{\mathrm{P}} - \ketbra{\mathrm{S}}{\mathrm{S}}$.  With these reduced versions, we can express the Hilbert space of interest as
\begin{equation}
    \mathcal{H} = \mathcal{H}_\mathrm{motion} \otimes \mathcal{H}_{\ell}^\prime \otimes \mathcal{H}_\Sun^\prime \otimes \mathcal{H}_\mathrm{atom}^\prime.
\end{equation}

Following \cite{Mitchison2016Realising}, we first consider that the cooling scheme is designed to ensure that the only way that the atom can be excited to $\ket{\mathrm{D}}$ is through the simultaneous annihilation of a phonon and a laser photon, the operator for which is $a_\mathrm{motion} a_\ell$.  The cooling cycle, then, gives rise to an interaction Hamiltonian of the form
\begin{align}
    V \propto \left( a_\mathrm{motion}\, a_\ell \,a_\Sun \,\sigma_+^{\mathrm{(atom)}} + a_\mathrm{motion}^\dagger \;a_\ell^\dagger\; a_\Sun^\dagger\; \sigma_-^{\mathrm{(atom)}} \right)\label{eq:InteractionV1}
\end{align}
where $\sigma_+^{\mathrm{(atom)}} \equiv \ketbra{\mathrm{P}}{\mathrm{S}}$ is the atomic raising operator.

We can now identify the system that exchanges energy with the motion by defining the virtual qubit raising operator
\begin{align}
    \sigma_+^{(\mathrm{V})} \equiv a_\ell\, a_\Sun \,\sigma_+^{\mathrm{(atom)}} \label{eq:VirtualRaising}
\end{align}
to cast (\ref{eq:InteractionV1}) in the form
\begin{align}
    V \propto \left(a_\mathrm{motion}\, \sigma_+^{\mathrm{(V)}} + a_\mathrm{motion}^\dagger \;\sigma_-^{\mathrm{(V)}} \right) \label{eq:InteractionV2}
\end{align}
The system now evolves in a smaller Hilbert space 
\begin{equation}
    \mathcal{H}' \equiv \mathcal{H}_\mathrm{motion} \otimes \mathcal{H}_\mathrm{V},
\end{equation}
where the virtual qubit subspace is only two dimensional. The virtual qubit described by raising operator (\ref{eq:VirtualRaising}) has an energy splitting at frequency
\begin{align}
    \omega_\mathrm{V} = & \,\, -\omega_\ell - \omega_2 + \omega_3 \nonumber \\
    =&\,\, \omega_\mathrm{motion}.
\end{align}

To find the temperature of the virtual qubit, we apply the statistics of thermal equilibrium between the populations in the excited state, $p_\mathrm{e}^{(i)}$, and the ground state, $p_\mathrm{g}^{(i)}$, to each 2-level sub system \cite{virtual},
\begin{equation}
    \frac{p_\mathrm{e}^{(i)}}{p_\mathrm{g}^{(i)}} = \exp \left( -\frac{\hbar \omega_i}{k_\mathrm{B} T_i}\right)
\end{equation}
where $k_\mathrm{B}$ is the Boltzmann constant.  We find
\begin{align}
    \exp \left( -\frac{\hbar \omega_\mathrm{V}}{k_\mathrm{B} T_\mathrm{V}}\right) = & \,\, \frac{p_\mathrm{e}^{(\mathrm{V})}}{p_\mathrm{g}^{(\mathrm{V})}} \nonumber \\
    = & \,\,  \frac{ p_\mathrm{g}^{(\Sun)} p_\mathrm{g}^{(\ell)} p_\mathrm{e}^{(\mathrm{atom})}  }{ p_\mathrm{e}^{(\Sun)} p_\mathrm{e}^{(\ell)} p_\mathrm{g}^{(\mathrm{atom})} } \nonumber \\
    = & \,\, \exp \left( \frac{\hbar \omega_2}{k_\mathrm{B} T_2} + \frac{\hbar \omega_\ell}{k_\mathrm{B} T_\ell} - \frac{\hbar \omega_3}{k_\mathrm{B} T_3} \right). \label{eq:expVirt}
\end{align}
Solving (\ref{eq:expVirt}) for $T_\mathrm{V}$ gives
\begin{align}
    T_\mathrm{V} = \frac{\omega_\mathrm{V}}{ \frac{\omega_3}{T_3} - \frac{\omega_2}{T_2} - \frac{\omega_\ell}{T_\ell} }.
\end{align}

\section{Virtual temperature in continuous sunlight}\label{appendix:AllThermal}
If we neglect the coupling to room-temperature black-body radiation and assume, in the service of providing a conservative estimate of the achievable temperature, that both of the non-laser-addressed transitions can be driven by sunlight at $T_\Sun$, we have
\begin{align}
    T_\mathrm{V} = \frac{\omega_\mathrm{V}}{ \frac{\omega_3}{T_\Sun} - \frac{\omega_2}{T_\Sun} - \frac{\omega_\ell}{T_\ell} }.
\end{align}
Again taking the limit as $T_\ell \rightarrow \infty$ yields
\begin{align}
    T_\mathrm{V} = & \,\,\frac{\omega_\mathrm{motion}}{\omega_3 - \omega_2} T_\Sun \nonumber \\
    = & \,\, \frac{\omega_\mathrm{motion}}{\omega_1} T_\Sun.\label{eq:AllThermalTV}
\end{align}
For alkaline earth ions, $k_\mathrm{B} T_\Sun < \hbar \omega_1$ and we see that even in this case, we expect most of the population to be in the motional ground state for a motional temperature given by (\ref{eq:AllThermalTV}).

\section{Spectral Radiance of Single Modes}\label{appendix:SRSM}
While equation (\ref{eq:BSAomega}) provides an explanation for how to connect the power spectral density of thermal light confined to q1D to the spectral radiance of black-body radiation in 3D, the spectral radiance of thermal light emerging from a single-mode fiber will be highly anisotropic.  In order to predict the energy density at a particular position in space in the far field, the angular distribution is needed, and this depends upon the fiber's mode area.

For a single, gaussian mode of radiation ($1/e$ field radius $w_0$) there are multiple mode areas that could be assigned.  For example, it is common to adopt the integrated intensity, or ``top hat'' definition \cite{siegman86}, $A_\mathrm{TH} \equiv P/I_\mathrm{max} = \int \mathrm{d} A \exp(-2\rho^2/w_0^2) = \frac{\pi}{2}w_0^2$ where $P$ is the power in the travelling-wave mode and $I_\mathrm{max}$ is the peak intensity at the center of the mode.  This is attractive from a radiometry perspective since it is the area of a hole in an opaque screen that would pass the same power $P$ from normally-incident plane waves of intensity $I_\mathrm{max}$.  Adopting this, with a straightforward application of paraxial gaussian optics we can write the the angular distribution of the spectral radiance in the form
\begin{equation}
    B(\omega, \theta) = \frac{S(\omega)}{A_\mathrm{TH}} \frac{2}{\pi}\left(\frac{\omega w_0}{2c} \right)^2  \exp\left(-2 \sin^2 (\theta) /\left( \frac{2c}{\omega w_0}\right)^2  \right).
\end{equation}
However, care must be used when interpreting this in the context of radiative thermal transport, as this would imply that the differential spectral radiance evaluated at the \emph{peak} of the angular distribution, $\theta = 0$, exceeds the value of a Planckian black body by a factor of 4:
\begin{align}
    B_\mathrm{max}(\omega) \mathrm{d} \Omega =& \,\,  \frac{S(\omega)}{A_\mathrm{TH}} \frac{2}{\pi}\left(\frac{\omega w_0}{2c} \right)^2 \mathrm{d} \Omega \nonumber \\
    =& \,\, S(\omega) \frac{4}{\lambda^2} \mathrm{d} \Omega \nonumber \\
    = & \,\, 4 B_\mathrm{P}(\omega) \mathrm{d} \Omega.\label{eq:4BP}
\end{align}

At first glance, this seems to violate thermodynamic principles.  For example, one could imagine the use of a series of fibers that are all carrying thermal radiation from a source at temperature $T$ to tile the full solid angle surrounding another body, thereby illuminating it with an approximately isotropic, average spectral radiance that is four times more powerful than that inside the source, which would allow it to equilibrate to a temperature exceeding the source.

However, while the mode area $A_\mathrm{TH}$ can be useful for describing the spatial distribution of power in a gaussian mode, the mode itself technically spans an infinite transverse extent, and this infinite support precludes the tiling of space by adjacent, orthogonal modes.  If we instead sharply cut off the Gaussian spatial mode at finite radius $R_\mathrm{c}$ to allow adjacent modes to be spaced by $2R_\mathrm{c}$, the degradation in peak spectral radiance per mode caused by spreading of the angular distribution from diffraction at the cutoff must to be taken into account.  For fixed total transmitted power spectral density per mode, the optimum cutoff radius is zero, asymptotically approaching a top-hat mode, for which the peak spectral density is a factor of 4 smaller than Eq.~(\ref{eq:4BP}).  It may, therefore, be safest to use a well-defined, finite support when defining the mode area for radiative thermal transport with gaussian modes, as the mode itself requires a larger area than just its variance (or full-width-at-half-maximum) to retain the far-field behavior described by gaussian optics.

The broader conclusion, here, is that care must be used when trying to use spectral radiance for single, isolated modes, as the mode area (analogous to position) and the mode solid angle (analogous to momentum) of a single mode cannot be sharply defined simultaneously.  A single-mode fiber emitting thermal radiation at temperature $T$ is in many ways similar to a black body at $T$, but its emitted radiance is not isotropic, and it does not follow the Lambert or Stefan-Boltzmann laws.  Many of the results that may be familiar for three-dimensional Planckian black-body radiation are not necessarily valid for an isolated spatial mode.

\end{appendices}
\end{document}